\documentclass[aps,prb,twocolumn]{revtex4}
\usepackage{epsfig}
\newcommand{\Sp}{\mbox{{\bf s}}}

\newcommand{\be}{\begin{equation}}
\newcommand{\ee}{\end{equation}}
\newcommand{\ben}{\begin{eqnarray}}
\newcommand{\een}{\end{eqnarray}}
\newcommand{\ra}{\rangle}
\newcommand{\la}{\langle}

\newcommand{\Neel}{N\'{e}el }

\newcommand{\psbild}[1]{#1}  % Linux Version, d.h. mit eps Bildern
\usepackage{epsfig}
\begin{document}

\title{
Direct calculation of the spin stiffness on square, triangular and cubic
lattices using the coupled cluster method}

\author{
S.E. Kr\"uger$^{\dagger}$, R. Darradi$^{\ddagger}$ and J. Richter$^{\ddagger}$
}
\affiliation
{
$^{\ddagger}$ Institut f\"ur Theoretische Physik, Universit\"at
Magdeburg, 
P.O.B. 4120, 39016 Magdeburg, Germany\\
$^{\dagger}$ IESK Kognitive Systeme, Universit\"at Magdeburg, PF 4120,  39016  Magdeburg, Germany
}
\author{D. J. J. Farnell}
\affiliation
{
Unit Of Ophthalmology, Department of Medicine,
University Clinical Departments, Daulby Street,
University of Liverpool, Liverpool L69 3GA, United Kingdom
}

\date{\today}

\begin{abstract}
We present a method for the direct calculation of the spin stiffness by
means of the coupled cluster method.  
For the spin-half Heisenberg 
antiferromagnet on the square, the triangular and the cubic 
lattices we calculate the stiffness 
in high orders of approximation.
For the square and the cubic lattices our results  are in very 
good agreement with the best results available in the literature.
For the triangular lattice our result is more precise than any other result 
obtained so far by other approximate method. 
\end{abstract}
  %Quantum spin model, magnetic phase boundaries, 
   %General theory and models of magnetic ordering

%]  % zu \twocolumn[... (alles zweispaltig ausser Abstrakt,
                       % NICHT bei preprint benutzen!)
\maketitle

%%%%%%%%%%%%%
\section{Introduction}
The study of quantum magnetism has attracted much
experimental and theoretical attention over many years, for an overview, see
Ref.~\onlinecite{book}.
The spin stiffness $\rho_s$ constitutes, together with the spin-wave velocity, 
a fundamental parameter that determines the low-energy dynamics of 
magnetic systems.\cite{halperin69,chakra89,chub94}
In particular, in two-dimensional  quantum antiferromagnets, where
magnetically ordered as well as quantum disordered ground-state  phases are
observed, the ground-state stiffness measures the distance of the 
ground state from criticality\cite{chub94} and can 
be used, in addition to the sublattice magnetization $M$, to test 
the existence or absence of magnetic long-range order (LRO). 

Over the last 15 years in a series of papers several methods
like series expansion,\cite{singh89,hamer94}
 spin-wave theory,\cite{igarashi91,hamer94,
chubukov94,lecheminant95,oles02,kr05} quantum Monte Carlo,\cite{makivic91}
exact diagonalization,\cite{bonca94,einarsson95,lecheminant95}
Schwinger-boson approach, \cite{auerbach88,croo97,manuel98}  and renormalization 
group theory\cite{spanu05} 
have been used 
to calculate the 
spin stiffness of the spin-half Heisenberg antiferromagnet
(HAFM) on the square, the triangular and  the cubic 
lattices.
However, results for the triangular lattice seem to be less precise than
those for the square lattice due to strong frustration. Published values 
therefore show significant variability.

The spin stiffness $\rho_s$ 
measures the increase in the amount of energy when we rotate the
order parameter of a magnetically long-range ordered system along a given
direction by a small  angle
$\theta$ per unit length, i.e. 
\be\label{stiffn}
\frac{E(\theta)}{N} = \frac{E(\theta=0)}{N} + \frac{1}{2}\rho_s \theta^2 +
{\cal O}(\theta^4)
\ee
where $E(\theta)$ is the ground-state energy as a function of the imposed
twist, and $N$ is the number of sites. In the thermodynamic limit, a positive
value of $\rho_s$ means that there is LRO in the system, while a value of zero 
reveals that there is no LRO.\cite{note} 

In this paper we present a new method to calculate the spin stiffness 
for the quantum-spin HAFM
using the 
coupled cluster approach.\cite{coester58,bishop87,bishop91}
The coupled cluster approach 
is a  powerful and universal tool in quantum many-body physics
which has been applied in various fields like nuclear physics, quantum
chemistry, strongly correlated electrons etc..\cite{coester58,bishop87,bishop91}
More recently the coupled cluster method (CCM) has  been
applied to quantum spin systems with much
success, see, e.g., Refs.~\onlinecite{rog_her90,bishop91a,bishop94,
zeng98,bishop98,krueger00,bishop00,krueger01,farnell01,kr01,rachid04,rachid05}.
In the field of magnetism an important advantage of this approach 
is its applicability to strongly frustrated quantum spin
systems in any dimension, 
where some other methods, such as, e.g., the quantum Monte Carlo method  
fail. Therefore the method to calculate the spin stiffness described in this
paper is quite generally applicable to spin systems also with non-collinear
ground states. 

To demonstrate the potential of the presented method we calculate the
spin stiffness 
for the spin-$\frac{1}{2}$ HAFM with nearest-neighbor 
interaction on the cubic, the square, and on the triangular lattices and
compare our results with 
available data in the literature.
While for the square and the cubic lattices accurate high order spin-wave
results are available which can be used to estimate the accuracy of the CCM
results,  the known results for the frustrated HAFM on
the triangular lattice with a non-collinear ground state seem to be less
reliable, since the used methods are less accurate. 
We argue
that our result for the 
stiffness of the HAFM
on the triangular lattice obtained by CCM in high order of approximation
is better than the so far available results. 
We mention that 
some preliminary results for the spin stiffness of the so-called $J-J'$ model
using the CCM can be found in Ref. \onlinecite{kr01}. 

\section{The method}
The model we consider is the spin-half HAFM 
\be \label{ham}
      H  =  J \sum_{\langle i,j\rangle }\Sp_i\cdot\Sp_j  . \ee
In (\ref{ham}) the sum runs over all pairs of nearest neighbors 
$\langle i,j\rangle$. We now set $J=1$ henceforth.

We start with  a brief illustration of the main features of the CCM.  For a
general  overview on the CCM the interested reader is referred, e.g., to
Ref.~\onlinecite{bishop91} and  for details of the CCM computational algorithm 
for quantum spin systems (with
spin quantum number $s=1/2$) to  Refs.~\onlinecite{zeng98,bishop00,krueger00}.
The starting point for a CCM calculation
is the choice of a normalized model or reference state $|\Phi\rangle$,
together with a set of mutually commuting
multispin creation operators $C_I^+$ which are defined over a complete set of
many-body configurations $I$. The operators
$C_I$ are the multispin destruction operators
and are defined to be the Hermitian
adjoints of the $C_I^+$. We choose $\{|\Phi\rangle;C_I^+\}$ in such a way
that we have $\langle\Phi|C_I^+=0=C_I|\Phi\rangle$, $\forall I\neq 0$.
Note that the CCM formalism corresponds to the thermodynamic limit $N\rightarrow\infty$.

For spin systems, an appropriate choice for the CCM model state $|\Phi\rangle$
is often a classical spin state, in which the most general
situation is one in which each spin can point in an arbitrary direction.

\begin{figure}[ht]
  \psbild{\centerline{\epsfysize=2.15cm \epsfbox{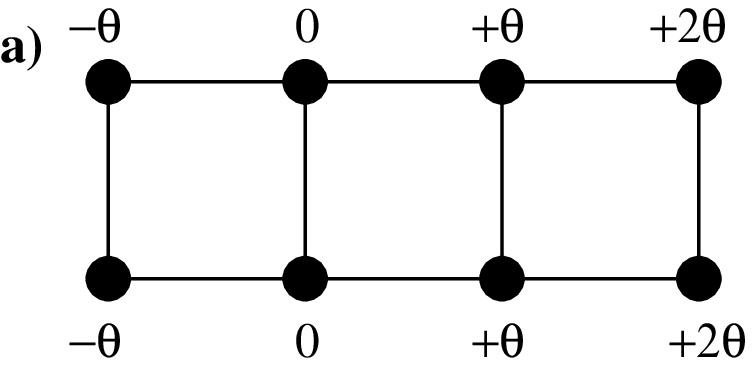}}
    \vspace{0.3cm}
    \centerline{\hspace{-0.9cm}\epsfysize=2.45cm \epsfbox{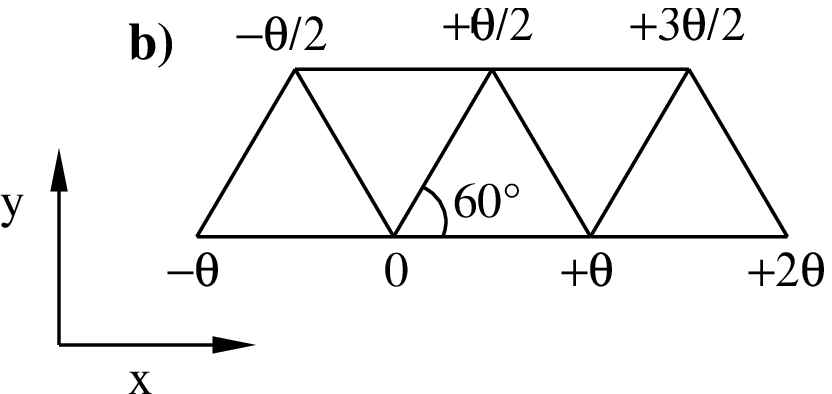}}}  
    \vspace{0.3cm}
  \caption{Illustration of the twisted \Neel state (a: square lattice; b:
       triangular lattice). The twist is introduced along rows in 
          $x$ direction. The angles at the lattice sites indicated the
        twist of the spins with respect to the corresponding \Neel state. }
  \label{fig_sti_bild}
\end{figure}

We then perform a local coordinate transformation such that all spins are 
aligned in negative $z$-direction in the 
new coordinate frame.\cite{krueger00,rachid05} As a result we have
\be 
   |\Phi\rangle=|\cdots\downarrow\downarrow\downarrow\cdots\rangle; \quad C_I^+=s_i^+,\,\,
s_i^+s_{j}^+,\,\, s_i^+s_{j}^+s_{k}^+,\cdots
,\ee
(where the indices $i,j,k,\dots\;$ denote arbitrary lattice sites) for the model state and the 
multispin creation operators which
now consist of spin-raising operators only.
In the new coordinate system the Hamiltonian reads \cite{krueger00} 
\ben
H&=&J \sum_{\langle i,j\rangle } \Bigg\{ 
\frac{1}{2}\sin\varphi[s_i^+s_j^z-s_i^zs_j^++s_i^-s_j^z-s_i^zs_j^-] \nonumber \\ 
   & &+\cos\varphi s_i^zs_j^z  
   +\frac{1}{4}(\cos\varphi+1)[s_i^+s_j^-+s_i^-s_j^+] \nonumber \\
   & &  +\frac{1}{4}(\cos\varphi-1)[s_i^+s_j^++s_i^-s_j^-]\Bigg\}
,\een
with $\varphi$ being the angle between the two spins, and
$s^{\pm}\equiv s^x\pm {\rm i}s^y$ are the spin-raising 
and spin-lowering operators.
According to Fig.~\ref{fig_sti_bild} we have e.g. for 
the twisted \Neel state on the square lattice $\varphi=\pi$ for
nearest-neighbors along the
$y$ direction but $\varphi=\pi+ \theta$ along the
$x$ direction and for the twisted $120^\circ$ \Neel state on the 
triangular lattice we have $\varphi=2\pi/3+ \theta/2$ for nearest neighbors
along the
$\frac{1}{2}\vec{e}_x+\vec{e}_y$ direction
but $\varphi=4\pi/3+ \theta$ along the
$x$ direction.

The CCM parameterizations of the ket and bra ground
states are given by
\begin{eqnarray}
\label{ket}
H|\Psi\rangle = E|\Psi\rangle\,\, ; \qquad  \langle\tilde{\Psi}|H = E
\langle\tilde{\Psi}| \; \; ;\nonumber\\
|\Psi\rangle = e^S|\Phi\rangle\,\, ; \qquad S = \sum_{I \neq
0}{\cal S}_IC_I^+\; \; ; \nonumber\\
\langle\tilde{\Psi}| =  \langle\Phi|\tilde{S}e^{-S}\,\, ; 
\qquad \tilde{S} =
1 + \sum_{I \neq 0}\tilde{\cal S}_IC_I^- \;.
\end{eqnarray}
The correlation operators $S$ and $\tilde {S}$ contain the  correlation coefficients
${\cal S}_I$ and $\tilde {\cal S}_I$ that we must determine.
Using the Schr\"odinger equation, $H|\Psi\ra=E|\Psi\ra$, we can now write
the ground-state energy as $E=\la\Phi|e^{-S}He^S|\Phi\ra$ and 
the sublattice magnetization is given
by $ M = -1/N \sum_i^N \la\tilde\Psi|s_i^z|\Psi\ra$, where $s_i^z$ is
expressed in the transformed coordinate system.
To find the ket-state and bra-state correlation coefficients we
require that the expectation value $\bar H=\langle\tilde\Psi|H|\Psi\rangle$
is a minimum with respect to the
bra-state and ket-state correlation coefficients, such that 
the CCM ket- and bra-state equations are given by
\begin{eqnarray}
\label{ket_eq}
\langle\Phi|C_I^-e^{-S}He^S|\Phi\rangle = 0 \qquad \forall I\neq
0\\
\label{bra_eq}
\langle\Phi|\tilde{\cal S}e^{-S}[H, C_I^+]e^S|\Phi\rangle = 0 \qquad \forall
I\neq 0.
\end{eqnarray}
The problem of determining the CCM equations now becomes a
{\it pattern-matching exercise} of the $\{C_I^-\}$ to
the terms in $e^{-S}He^S$ in Eq.~(\ref{ket_eq}). 

The CCM formalism is exact if we take into account all possible multispin
configurations in
the correlation operators $S$ and $\tilde S$, This is, however, generally not 
possible for most quantum many-body models including those studied here. 
We must therefore use the most common  
approximation
scheme to truncate the expansion
of $S$ and $\tilde S$ in the Eqs.~(\ref{ket_eq}) and (\ref{bra_eq}), namely
the LSUB$n$ scheme,
where we include only $n$ or fewer correlated
spins in all configurations
(or lattice animals in the language of graph theory)
which span a range of no more than $n$ adjacent (contiguous)
lattice sites 
(for more details see Refs. \onlinecite{bishop91a,krueger00,bishop00}).

The spin stiffness considered in this paper is the stiffness of the \Neel
order parameter (sublattice magnetization). Hence the corresponding 
model state $|\Phi\rangle$ is the \Neel state. This is the ordinary
collinear two-sublattice \Neel state for the square and the cubic lattices. 
The model state is a noncollinear $120^\circ$ three-sublattice \Neel
state  for the triangular lattice. Note that for the
collinear \Neel state only LSUB$n$ approximations with even $n$ are
relevant.\cite{bishop00,krueger00}    
In order to calculate the spin stiffness directly using Eq.~(\ref{stiffn}) 
we must modify the model (N\'{e}el) state by introducing an appropriate twist
$\theta$, see Fig.~\ref{fig_sti_bild}.
Thus 
the ket-state correlation coefficients ${\cal S}_I$ (after solving the
CCM equations (\ref{ket_eq})) depend on $\theta$ and hence the ground-state 
energy $E$ is also dependent on $\theta$. 
Note that our numerical code for 
the CCM-LSUB$n$ approximation allows us to calculate $E(\theta)$
with very high precision of about 14 digits.
First we have checked numerically 
that the ground-state  energy calculated in LSUB$n$ approximation 
does indeed fulfill the relation (\ref{stiffn}) with high precision for 
$\theta \lesssim 0.01$.
The stiffness now can easily be
calculated using numerical differentiation of $E(\theta)$ which was done 
using a three-point formula with
$\theta=-10^{-4},0,+10^{-4}$. 

\begin{figure}[ht]
\psbild{\centerline{\epsfysize=6.5cm \epsfbox{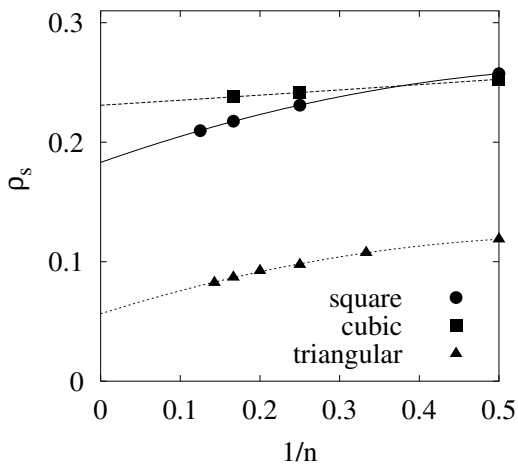}}}   
\caption{Extrapolation of the CCM-LSUB$n$ results for the stiffness. The
points represent the CCM-LUB$n$ results and the lines correspond to the 
function (\ref{scal}) fitted to this data points.}
\label{fig_extr}
\end{figure}

Since the LSUB$n$ 
approximation becomes exact for   $n \to \infty$, it 
is useful to extrapolate the 'raw' LSUB$n$
results to the limit $n \to \infty$. 
Although we do not know the  exact scaling of the LSUB$n$ results, 
there is some empirical experience\cite{krueger00,zeng98,bishop00} 
how the ground-state energy and the order parameter 
for antiferromagnetic spin models scale with $n$.
Based on this experience we have tested 
several fitting functions 
for the stiffness and we have found the best 
extrapolation is obtained by the
fitting function   
\be \label{scal}
a=a_0+a_1 \frac{1}{n}+a_2\frac{1}{n^2} . 
\ee
This law is
known\cite{bishop00,krueger00,krueger01,farnell01,rachid04,rachid05}
to provide good extrapolated results for the order 
parameter.
We show  this extrapolation in
Fig.~\ref{fig_extr}. 

\section {Results}
Let us start with the results for the square lattice. Exploiting  the
lattice symmetries we are able to perform calculations up to LSUB8, where
for the twisted state 21124 ket equations (\ref{ket_eq}) have to be solved.  
The results for the stiffness are given in Tab.~\ref{tab1}. 
\begin{table}[ht]
\caption{\label{tab1}Spin stiffness $\rho_s$ 
for the spin-half Heisenberg 
antiferromagnet on the square lattice
calculated by various CCM-LSUB$n$ approximations and the result of the 
$n \to \infty$ extrapolation using LSUB$n$ with $n=4,6,8$.}
\begin{ruledtabular}
\begin{tabular}{rr|r}
LSUB$n$ & number eqs. & $\rho_s$\\
\hline
2 &    3     &   0.2574  \\
4 &   40     &   0.2310  \\
6 &   828     &   0.2176  \\
8 &    21124 &  0.2097  \\
extrapol. & -- & 0.1812\\
\end{tabular}
\end{ruledtabular}
\end{table}
\begin{table}[ht]
\caption{\label{tab2}Collection of data for the spin stiffness 
$\rho_{s}$ for the spin-half Heisenberg 
antiferromagnet on the square lattice calculated by different methods.}
\begin{ruledtabular}
\begin{tabular}{c|c}
first-order spin-wave theory\cite{igarashi91,hamer94}     &   0.191  \\
second-order spin-wave theory\cite{igarashi91,hamer94}   &   0.181\\
third-order spin-wave theory\cite{hamer94} &   0.175  \\
series expansion\cite{hamer94}&  0.182  \\
exact diagonalization\cite{einarsson95}& 0.183 \\
quantum Monte Carlo\cite{makivic91}& 0.199 \\
Schwinger-boson approach I\cite{auerbach88}& 0.176 \\
Schwinger-boson approach II\cite{manuel98}& 0.153 \\
CCM& 0.181 \\
\end{tabular}
\end{ruledtabular}
\end{table}
\begin{table}[ht]
\caption{\label{tab3}In-plane spin stiffness $\rho_{s}$ 
for the spin-half Heisenberg 
antiferromagnet on the triangular lattice
calculated by various CCM-LSUB$n$ approximations and the result of the 
$n \to \infty$ extrapolation using LSUB$n$ with $n=2,3,4,5,6,7$.}
\begin{ruledtabular}
\begin{tabular}{rr|r}
LSUB$n$ & number eqs. & $\rho_{s}$\\
\hline
2 &     3 & 0.1188\\
3 &    14 & 0.1075\\
4 &    67 & 0.0975\\
5 &   370 & 0.0924\\
6 &  2133 & 0.0869\\
7 & 12878 & 0.0824\\
extrapol. & -- & 0.0564\\
\end{tabular}
\end{ruledtabular}
\end{table}
Using LSUB$n$ with $n=2, 4, 6, 8$ the extrapolated result is $\rho_s=0.1831$.
As known from the sublattice magnetization even better results can be
obtained by excluding the LSUB2 data. 
Indeed the extrapolation using the LSUB4, LSUB6, LSUB8 data yields 
$\rho_s=0.1812$.
Note that the
corresponding extrapolated value for the sublattice magnetization 
\cite{bishop00} $M=0.3114$
is in good agreement with other results\cite{bishop00,richter04}. 
A certain estimate of the accuracy can be obtained 
by an extrapolation using
LSUB2, LSUB4, LSUB6, only, which yields  $\rho_s=0.1839$.
We compare our results for $\rho_s$ with some data obtained  
by other methods in Tab.~\ref{tab2}.
Obviously, there is a significant variance in the data. In particular, 
the value obtained by quantum Monte Carlo seems to be surprisingly large. 
However, this might be connected with the fact, that in 
Ref. \onlinecite{makivic91} the stiffness was not determined directly, 
but via the temperature dependence of the correlation length 
which may lead to larger uncertainty.     
We think, that the high-order spin-wave theory\cite{hamer94}
is the most systematic approach, since one can see how the stiffness 
changes with increasing order of approximation. Assuming the 
third-order order spin-wave results as a benchmark we find that our CCM result 
deviates by about 3\%. 

%%%%%%%%%%%%%%%%%%%%%%%%%%%%%%%%%%%%%%%%%%%%%%%%%%%%%%%%%
\begin{table}[ht]
\caption{\label{tab4}Collection of data for the spin stiffness
$\rho_{s}$ for the spin-half Heisenberg 
antiferromagnet on the triangular lattice calculated by different methods.}
\begin{ruledtabular}
\begin{tabular}{c|c}
exact diagonalization\cite{lecheminant95}& 0.05 \\
first-order spin-wave theory\cite{chubukov94,lecheminant95}& 0.080 \\
Schwinger-boson approach\cite{manuel98}& 0.088 \\
CCM& 0.056 \\
\end{tabular}
\end{ruledtabular}
\end{table}
%%%%%%%%%%%%
\begin{table}[ht]
\caption{\label{tab5}
Spin stiffness $\rho_s$ 
for the spin-half Heisenberg 
antiferromagnet on the cubic lattice
calculated by various CCM-LSUB$n$ approximations and the result of the 
$n \to \infty$ extrapolation using LSUB$n$ with $n=2,4,6$.}
\begin{ruledtabular}
\begin{tabular}{rr|r}
LSUB$n$ & number eqs. & $\rho_{s}$\\
\hline
2 &    4 & 0.2527\\
4 &   106 & 0.2416\\
6 & 5706& 0.2380\\
extrapol. & -- & 0.2312\\
\end{tabular}
\end{ruledtabular}
\end{table}

For the triangular lattice the twist we consider (see Fig.~\ref{fig_sti_bild})
corresponds to the in-plane spin stiffness. 
Due to the noncollinear structure of
the three-sublattice \Neel state also LSUB$n$ approximations
with odd $n$ appear. Furthermore the 
number of ket equations in a certain level of approximation becomes 
larger then for the square lattice and as a results 
the highest
level of approximation we are able to consider is LSUB7.  
The results for different LSUB$n$ approximations 
are given in Tab.~\ref{tab3}.   
The extrapolation of the LSUB$n$ data according to Eq.~(\ref{scal})
with $n=2, 4, 6$ leads to
$\rho_{s}=0.0604$
and with $n=2, 3, 4, 5, 6, 7$ to
$\rho_{s}=0.0564$. Again the difference in the two values 
can be considered as a certain estimate of the accuracy. 
As a byproduct of our high-order calculation 
we can give here improved values for 
the sublattice magnetization $M$.
So far results for $M$ up to LSUB6\cite{zeng98,farnell01} are
published. We can add  $M=0.3152$ (LSUB7) and $M=0.3018$ 
(LSUB8).  
The corresponding extrapolated value using Eq..~(\ref{scal}) 
and  LSUB$n$ with $n=2,3,4,5,6,7,8$
is $M=0.2134$, which  
is close to spin-wave\cite{chubukov94,miyake} and Green's function Monte
Carlo\cite{capriotti99} results. 
The small values of the stiffness and the order parameter in comparison with
the 
square lattice are attributed to the frustration leading to a noncollinear
ground state and in combination with quantum fluctuations to a drastic 
weakening of magnetic order in the spin-half HAFM.

We compare our results for $\rho_s$  with available results from
literature, see Tab.~\ref{tab4}. Comparing the methods used to calculate
$\rho_s$ for the square
lattice (Tab.~\ref{tab2}) and for the triangular lattice (Tab.~\ref{tab4})
we see that the results for the triangular lattice are much less
reliable, since here the accuracy of the methods used in
Refs.~\onlinecite{chubukov94,lecheminant95,manuel98} is limited.
Assuming the same tendency as for the square lattice
we can expect that the
first-order spin-wave value for $\rho_s$ 
\cite{chubukov94,lecheminant95}  becomes smaller (and therefore closer to
our CCM result) going to second- and third-order spin-wave theories.
We believe that our result is indeed of higher accuracy than data for
$\rho_s$
so far available.

We present now our results for $\rho_s$  for the simple cubic lattice, see
Tab.~\ref{tab5}. Here the highest
level of approximation we can consider is LSUB6.
From Fig.~\ref{fig_extr} it becomes obvious, that there is only a weak 
dependence on the level of CCM approximation $n$. 
Therefore we expect that the extrapolation according to Eq.~(\ref{scal}) 
yielding 
$\rho_s=0.2312$ is particular accurate.
Indeed we find that our result
is in very good agreement 
with the result obtained by second-order spin-wave theory\cite{hamer94}
$\rho_s=0.2343$. Note that the $1/s$ spin-wave expansion seems to 
converge very rapidly\cite{hamer94} and therefore the second-order spin-wave 
theory is expected to yield a very  precise result for $\rho_s$.  
For the sublattice magnetization a corresponding
extrapolation leads to $M=0.4181$\cite{bishop00} coinciding to 1\% with the high precision
third-order spin-wave result.\cite{oitmaa94}

\section{Summary}
In summary, we have presented a method for the direct calculation of the
spin stiffness within the framework of the coupled cluster method.
We obtain accurate values for the stiffness by applying this algorithm 
to high orders of LSUB$n$ approximations for 
the spin-half isotropic Heisenberg 
antiferromagnet on various lattices with and without frustration. 

{\it   Acknowledgment:}
This work was supported by the DFG
(Ri615/12-1). The authors thank J. Schulenburg for assistance in numerical calculations.

%\renewcommand{\baselinestretch}{1.0}  % Faktor des Zeilenabstandes
%%%%%%%%%%%%%%%%%%%%%%%%%%%%%%%%%%%%%%%%%%%%
%\begin{references}

%%%%%%%%%%%%%%%%%%%%%%%%%%%%%%%%%%%%%%%%%%%%%%%%%%%%%%%%%%%%%%%%%%%%%%%%

\begin{thebibliography}{99}
\bibitem{note} The stiffness $ \rho_s$  can be calculated also for 
finite systems.
However, for finite systems one can get a nonzero value 
for $\rho_s$ even for models where in the thermodynamic limit 
the \Neel LRO breaks down, see Ref. \onlinecite{einarsson95}.
\end{thebibliography}
\end{document}